\documentstyle[12pt]{article}
\begin{document}
\newcommand{\ai}{\mbox{a}}
\newcommand{\bi}{\mbox{b}}
\newcommand{\at}{\mbox{$\tilde{\mbox{a}}$}}
\newcommand{\bt}{\mbox{$\tilde{\mbox{b}}$}}
\newcommand{\sai}{\mbox{\small a}}
\newcommand{\sbi}{\mbox{\small b}}
\newcommand{\sat}{\mbox{$\tilde{\mbox{\small a}}$}}
\newcommand{\sbt}{\mbox{$\tilde{\mbox{\small b}}$}}
\newcommand{\be}{\begin{equation}}
\newcommand{\ee}{\end{equation}}
\newcommand{\vth}{\vspace{3mm}}
\newcommand{\bec}{\begin{center}}
\newcommand{\ec}{\end{center}}
\newcommand{\befl}{\begin{flushleft}}
\newcommand{\efl}{\end{flushleft}}
\newcommand{\hed}{\qquad }
\begin{flushright}
TMUP-HEL-9808\\
\small{revised 3 November 1998}
\end{flushright}
\bec
\huge{ Two-dimensional Liouville Gravity Theory
             with Non-Trivial Classical Background }\\[1.0cm]
\large{Noriaki Ano}$~ ^\star$
\footnote[0]{$\star$ E-mail: ano@phys.metro-u.ac.jp}\\
\normalsize{\it Department of Physics,
Tokyo Metropolitan University,}\\
\normalsize{\it Minamiohsawa 1-1, Hachiohji, 192-03, Japan}\\
\large{Takashi Suzuki}$~ ^{\star\star}$
\footnote[0]{$\star\star$ E-mail: stakashi@cc.it-hiroshima.ac.jp}\\
\normalsize{\it Department of Electronics, Hiroshima Institute of 
Technology,}\\
\normalsize{\it Saeki, Hirosima 731-51, Japan}\\[1.0cm]
\ec
\normalsize
\bec {\large Abstract} \ec
We examine a possibility to introduce a non-trivial classical 
background metric into the 2-d gravity theory. 
The classical background appears as a nontrivial part
of the Wely scaling factor of 
physical metric of 2-d surfaces with some conformal dimension. 
On the other hand, the rest part of the factor corresponds to 
the quantum fluctuating sector, having another conformal dimension 
such that these two conformal dimensions 
sum up to just (1,1). 
We finally claim that in the 2-d gravity theory,
there should be possibility that
the target space dimensions $D$ go beyond 1. 

\newpage
\normalsize
\hed
What we know as non-critical string theory is the so-called 
two dimensional Liouville gravity introduced by Polyakov \cite{Po1} 
and developed by lots of authors (see, for example,  review  articles 
\cite{Se}). 
In particular, a family of recent works on the Liouville gravity
motivated by the works of Knizhnik-Polyakov-Zamolodchikov 
(KPZ) \cite{KPZ} and of David \cite{Da}-Distler-Kawai\cite{DK} 
(DDK) have brought a great progress in  the 2-d gravity theory.  
These  approaches to 2-d gravity via the Liouville theory including  
the matrix model  
treat only quantum fluctuation without any considerations of 
classical backgrounds, since these approaches are perfectly independent of 
background metric. 

Let us consider in more detail the above statement by     
focusing on the DDK approach with the Liouville action 
\be
S_L(\Phi;\hat{g})\,=\,\frac{1}{8\pi} \int d^2\xi \sqrt{\hat{g}}\left(
(\hat{\nabla} \Phi)^2 +  Q\, \hat{R}\,\Phi
\right), 
\label{Liouville}
\ee
where $\Phi(z,\bar{z})$ is the Liouville field and 
$\hat{g}$ is a background metric, i.e.,
the physical metric of 2-d
surface is given by $g=e^{\gamma\Phi}\,\hat{g}$.
Looking at the action (\ref{Liouville}), 
the DDK approach is just a conformal field theory of 
the Chodos-Thorn/Feigin-Fuks (CTFF) type \cite{CTFF}.
The conformal invariance of this theory is meant by
the following description:
The counter transformation
(conformal transformation)
$z\rightarrow f(z), \,\bar{z} \rightarrow \bar{f}(\bar{z})$ 
against the Weyl rescaling
$g \rightarrow e^{\sigma}g$
which we do treat equivalently as
$\hat{g} \rightarrow e^{\sigma}\hat{g}$ after all,
fulfills the condition
\be
\left\vert\,\frac{df}{dz}\,\right\vert^{\,2} =
e^{\sigma(z,\bar{z})},
\label{conftra}
\ee
that is, the theory is invariant under the
transformations
\begin{eqnarray}
\gamma\Phi(z,\bar{z}) \enskip & \longrightarrow & \enskip
\gamma\Phi(z,\bar{z}) - 
\log\,\left\vert\,\frac{df}{dz}\,\right\vert^{\,2}\,= \, 
\gamma\Phi(z,\bar{z})-\sigma(z,\bar{z}), \nonumber \\
\hat{g} & \longrightarrow & e^{\sigma} \hat{g}.
\label{first}
\end{eqnarray}
In this description,
it is contained that the conformal weights of $e^{\gamma \Phi}$
and $\hat{g}$ are (1, 1) and (0, 0), respectively.
We know that
this fact is crucial in the ordinary approach to
the Liouville gravity theory, in which
the quantum fluctuation of Liouville field
is independent of choice of both its origin and
the dimensionless background metric $\hat{g}$.
Therefore, there is no room in the DDK approach
to put into the theory any information
about the classical background but moduli,
as long as the conformal invariance is held
together with the fact that the Lioville field $\Phi$ is the
full quantum object of weight (1, 1).
In the well-known approach of DDK to the 2-d gravity theory,
it seems that
the way to assign the full conformal weight (1, 1) of
the Weyl scaling factor to quantum and classical sectors
of physical metric $g$ is unique.
Is there another significant way to assign the weight
to each sectors of $g$ ?

It may be possible to introduce a different assignment
of weight (1, 1) from the ordinary manner of DDK 
into the 2-d gravity theory
so as not to spoil the conformal invariance.
The motivation of this short paper is to find
this possibility, and we will present another approach 
to the Liouville gravity theory
by introcucing nontrivial classical background.
The classical background, which will be introduced in this paper, 
appears as a nontrivial part of the
Weyl scaling factor $e^{\Phi}$ of physical metric $g$. 
The importance of the introduction of such a classical sector 
has been pointed out via quantum group considerations
in Ref.\cite{TS}. \\

A key in our approach is the following decomposition of the 
Weyl scaling factor,
\be
e^{\Phi} = e^{\phi_{cl}} e^{\phi_{q}}
\ee
where $\phi_{q}$ is a quantum fields on which the path integration will 
be performed, whereas $\phi_{cl}$ is a classical field. 
That is, the quantum fluctuations of a 2-d surface
of physical metric $g$ must be measured
not around $\hat{g}$ but around $e^{\phi_{cl}}\hat{g}$. 
The essential point here is that
unlike the DDK approach in which 
the whole Weyl factor $e^{\Phi}$ is
the quantum object of weight (1,1) 
and $\hat{g}$ is of no dimension,
the classical background in our approach has 
a non-zero conformal dimension,
and it should also be emphasized that we require 
this classical-quantum composite system to be conformally invariant. 
We can see that in this case,
the reference metric $\hat{g}$ turns out to
be a part of the classical background.

We then obtain the following fundamental relation;
\be
\Delta_{\Phi} = \Delta_{cl} + \Delta_{q} =1,
\label{dimension-1}
\ee
where $\Delta_{\Phi}$, $\Delta_{cl}$ and $\Delta_{q}$ 
are the holomorphic conformal dimensions of $e^{\Phi}$, 
$e^{\phi_{cl}}$ and $e^{\phi_{q}}$, respectively. 
Therefore, this classical-quantum
composite system leads us to
the fact that the Weyl rescaling
$\hat{g} \rightarrow e^{\sigma}\hat{g}$
must be absorbed through the conforml transformations
of both classical $e^{\phi_{cl}}$
and quantum $e^{\phi_{q}}$ sectors,
which share the whole weight (1,1) of $e^{\Phi}$
with each other.
One must not mean in a simplistic way an origin of the
whole quantum factor $e^{\Phi}$ of
dimension (1,1) by the classical factor $e^{\phi_{cl}}$ in this paper.
We should take into account
not only the quantum fluctuating 
sector $e^{\phi_q}$ but also the classical background
$e^{\phi_{cl}}\hat{g}$ of non-zero weight.
As will be seen,
the classical background of non-zero weight
have non-negligible  
effect on the 2-d gravity theory. \\

Now, we are at the stage to investigate explicitly the 2-d
gravity theory with the nontrivial classical background. 
Let us start with observing the conformal transformations
of the factors $e^{\phi_{cl}}$ and $e^{\phi_{q}}$
of weights $\Delta_{cl}$ and $\Delta_{q}$ respectively.
As in the DDK approach, we recognize
that the Weyl rescaling of $\hat{g}$;
\be
\hat{g}\,\,\longrightarrow\,\, e^{\sigma(z,\bar{z})}\,\hat{g}
\label{weyl}
\ee
must be absorbed by the conformal transformation 
$z\rightarrow f(z), \,\bar{z} \rightarrow \bar{f}(\bar{z})$ 
with the relation (\ref{conftra}),
which yields the shifts of
$\phi_{q}$ and $\phi_{cl}$ as follows;
\begin{eqnarray}
\phi_{q}(z,\bar{z}) & \longrightarrow &
\phi_{q}(z,\bar{z}) - 
\Delta_{q}\,\,\log\,\left\vert\,\frac{df}{dz}\,\right\vert^{\,2}\,= \,
\phi_{q}(z,\bar{z}) - 
\Delta_{q}\,\sigma(z,\bar{z}), \nonumber   \\[.15cm]                      
\phi_{cl}(z,\bar{z}) & \longrightarrow &
\phi_{cl}(z,\bar{z}) - 
\Delta_{cl}\,\log\,\left\vert\,\frac{df}{dz}\,\right\vert^{\,2}\,= \,
\phi_{cl}(z,\bar{z}) - 
\Delta_{cl}\,\sigma(z,\bar{z}).
\label{gogo}
\end{eqnarray}
The invariance under the transformations
(\ref{weyl}) and (\ref{gogo}) means that
the theory does not depend on the choice of
$\hat{g}$ as the gauge fixing condition,
because the total conformal dimension of
$e^{\phi_{q}}$ and $e^{\phi_{cl}}$
is weight (1,1) of the whole Weyl scaling factor $e^{\Phi}$
as shown in the Eq.(\ref{dimension-1}).

Next, in order to present an explicit description of our model,
let us introduce actions for $\phi_{cl}$ and $\phi_{q}$ 
through the Jacobian $J(\phi_{cl}, \phi_{q}; \hat{g})$ 
of the path-integral measure under the replacement 
$g\,\rightarrow\, e^{\phi_{cl}}\ e^{\phi_{q}}\,\hat{g}$;
\be
{\cal D}_{g} \Phi {\cal D}_{g} (\mbox{gh}) {\cal D}_{g} X 
\, =\, {\cal D}_{\hat{g}} \phi_{q} {\cal D}_{\hat{g}} (\mbox{gh})
   {\cal D}_{\hat{g}} X J(\phi_{cl}, \phi_{q}; \hat{g}). 
\ee
We suppose that the Jacobian is given by the following formula;
\be
J(\phi_{cl}, \phi_{q}; \hat{g})=
    e^{-S[\phi_{cl};\hat{g}]-S[\phi_{q};\hat{g}]},
\ee
where $S[\phi_{cl};\hat{g}]$ and $S[\phi_{q};\hat{g}]$ are
the following Liouville-like actions;
\begin{eqnarray}
S[ \phi_{q};\hat{g}] & = & \frac{1}{8\pi}
  \int d^2 \xi \sqrt{\hat{g}}( \ai ( \hat{\nabla}\phi_{q} )^2
         + \bi \hat{R}\phi_{q}) \label{cl-liouville}\\
\vspace*{.3cm}
S[ \phi_{cl};\hat{g} ] & = & \frac{1}{8\pi}
  \int d^2 \xi \sqrt{\hat{g}}( \at ( \hat{\nabla}\phi_{cl} )^2
         + \bt \hat{R}\phi_{cl}).\label{q-liouville}
\end{eqnarray}
The cosmological constant term in each action is not considered 
for simplicity.\footnote{
Some differences from Ref.\cite{TS} appear here. 
For instance, in Ref.\cite{TS}, the background metric in $S[\phi_q]$
is different from one in Ref.\cite{TS}, etc.}

What we have to do next is to
determine the constants ``$\ai$,$\bi$,
$\at$ and $\bt$'' so that the total system, i.e.,
the $D$ string coordinates,
the reparametrization ghosts and the above $\phi_{cl},\,\phi_q$,
is invariant under the transformations (\ref{weyl})(\ref{gogo}).
For this purpose, it is enough to consider an infinitesimal 
transformations up to first order in $\sigma$,
and one finds the following terms;   \\

From $\phi_{q}$ sector: 
\begin{eqnarray}
&&\frac{\sai}{8\pi} \sqrt{\hat{g}} (\hat{\nabla} \phi_{q})^2\,\,
\rightarrow\,\, \frac{\sai}{8\pi} \sqrt{\hat{g}}(\hat{\nabla} \phi_{q})^2
              + \frac{\sai}{4\pi} \Delta_{q} \sqrt{\hat{g}} {\hat{\nabla}}^2
              \phi_{q} ~ \sigma,\nonumber \\[.2cm]
&&\frac{\sbi}{8\pi} \sqrt{\hat{g}} \hat{R} \phi_{q}\,\,
\rightarrow\,\, \frac{\sbi}{8\pi} \sqrt{\hat{g}} \hat{R} \phi_{q}
  - \frac{\sbi}{8\pi} \Delta_{q} \sqrt{\hat{g}} \hat{R} ~ \sigma 
- \frac{\sbi}{8\pi} \sqrt{\hat{g}}{\hat{\nabla}}^2 \phi_{q} ~ \sigma,
\nonumber
\end{eqnarray}

From $\phi_{cl}$ sector: 
\begin{eqnarray}
&&\frac{\sat}{8\pi} \sqrt{\hat{g}} (\hat{\nabla} \phi_{cl})^2\,\,
\rightarrow\,\, \frac{\sat}{8\pi} \sqrt{\hat{g}}(\hat{\nabla} \phi_{cl})^2
              + \frac{\sat}{4\pi} \Delta_{cl} \sqrt{\hat{g}} 
{\hat{\nabla}}^2
              \phi_{cl} ~ \sigma, \label{first-order}\\[.2cm]
&&\frac{\sbt}{8\pi} \sqrt{\hat{g}} \hat{R} \phi_{cl}\,\,
\rightarrow\,\, \frac{\sbt}{8\pi} \sqrt{\hat{g}} \hat{R} \phi_{cl}
  - \frac{\sbt}{8\pi} \Delta_{cl} \sqrt{\hat{g}} \hat{R} ~ \sigma 
  - \frac{\sbt}{8\pi} \sqrt{\hat{g}}{\hat{\nabla}}^2 \phi_{cl} ~ \sigma,
\nonumber
\end{eqnarray}

Finally, from path-integral measure:
\begin{displaymath}
\frac{D-25}{48\pi}\sqrt{\hat{g}}\hat{R}~ \sigma.
\end{displaymath}

\normalsize
As a consequence,
the requirement for the cancellations of these anomalous terms 
leads us to the condition;
\be
-(2 \Delta_{q} \ai - \bi ) \sqrt{\hat{g}}
{\hat{\nabla}}^2 \phi_{q} ~ \sigma
-(2 \Delta_{cl} \at -\bt ) \sqrt{\hat{g}}
{\hat{\nabla}}^2 \phi_{cl} ~ \sigma
+ \left(\frac{D-25}{6}+ \Delta_{q} \bi + \Delta_{cl} \bt\right)
\sqrt{\hat{g}} \hat{R} ~ \sigma = 0.\label{cft-condition-1}
\ee
We then obtain the following three equations;
\begin{eqnarray}
2 \Delta_{q} \ai - \bi  =  0, \label{three-eq-1}\\
2 \Delta_{cl} \at - \bt  =  0,\label{three-eq-2}\\
\frac{D-25}{6} + \Delta_{q} \bi + \Delta_{cl} \bt  =  0.\label{three-eq-3}
\end{eqnarray}
From eq.(\ref{three-eq-1})--(\ref{three-eq-3}), 
we have the condition for $D$;
\be
\frac{D-25}{12} + (\Delta_{q})^2 \ai +  (\Delta_{cl})^2 \at = 0.
\label{cft-condition-2}
\ee
It is easy to check that the second order calculation for $\sigma$
induces the same equation as (\ref{cft-condition-2}).
As a matter of course, the central charge of the whole theory
must be canceled out to zero, as long as we require
the conformal invariance on the whole.
We can see that eq.(\ref{cft-condition-2}) is
nothing but the condition that  the central charge of the 
total system vanishes.

Further, it is necessary to renormalize the quantum field 
$\phi_{q}$, so that we have the conventionally normalized kinetic 
term deriving the OPE $\phi_{q}(z)\phi_{q}(w)\sim -\ln(z-w)$.
Upon eq.(\ref{three-eq-1}) and the rescaling 
$\sqrt{\ai} \phi_{q} \rightarrow \phi_{q}$, 
the quantum action $S[\phi_{q};\hat{g}]$ reduces to 
\be
S[\phi_{q};\hat{g}]=\frac{1}{8\pi} \int d^2 \xi \sqrt{\hat{g}}
  \left( ( \hat{\nabla} \phi_{q} )^2 \phi_{q} 
   + 2 \Delta_{q} \sqrt{\ai} \hat{R}\phi_{q} \right),
\label{quantum-action}
\ee
and the Weyl factor $e^{\phi_{q}}$ must be modified as well;
\be
e^{\phi_{q}} \rightarrow e^{\alpha \phi_{q}}.
\ee
Here the new parameter $\alpha$ is introduced, since
the quantum part may include some quantum effect.

It is, however, possible to obtain the relation between 
\lq\lq a" and $\alpha$.   
To this end, we derive the energy-momentum tensor of 
the quantum sector $T_{q}$ from eq.(\ref{quantum-action}), 
\be
T_{q}= -\frac{1}{2} (\partial \phi_{q})^2 +
        \Delta_{q} \sqrt{\ai} \partial^2 \phi_{q}.
\ee
Then the conformal dimension $\Delta_{q}$
of the operator $e^{\alpha\phi_q}$
is calculated by the OPE,
$T_{q}(z)e^{\alpha\phi_{q}(w)}\sim
\Delta_{q}e^{\alpha \phi_{q}}(w)/(z-w)^2 
+ \cdot \cdot \cdot$,
and we obtain the equation
\be
\Delta_{q}=-\frac{\alpha}{2}(\alpha - 2 \Delta_{q} \sqrt{\ai}),
\label{delta-ope}
\ee
or, equivalently, 
\be
\sqrt{\ai}=\frac{1}{\alpha}+\frac{\alpha}{2\Delta_{q}}.
\label{sqrt-1}
\ee

Before going ahead, it is interesting to look at the central charge
of the theory.
The contribution of the quantum sector is calculated by the OPE,  
$T_{q}(z)T_{q}(w)$, which has the leading short distance behavior
$\frac{c_{q}}{2}/(z-w)^4 + \cdot \cdot\cdot$,
and it is $c_{q}=1 + 12(\Delta_{q}\sqrt{\ai})^2$.
On the other hand, one easily finds
the central charge coming from the
classical sector by imposing the conformal transformation,
as shown in the second line of (\ref{gogo}),
on the energy-momentum tensor of
the classical part, and by observing the anomalous term called
the Schwarizian derivative term.
We then find $c_{cl}= 12 (\Delta_{cl}\sqrt{\at})^2$.
Therefore, the total central charge amounts to 
\be
c^{tot}=c_X+c_{gh}+c_{cl}+c_q = D +(-26) + (1 + 12(\Delta_{q}\sqrt{\ai})^2)
+ 12 (\Delta_{cl}\sqrt{\at})^2,
\ee
and $c^{tot}/12$ is just the left hand
side of eq.(\ref{cft-condition-2}).\\

Now we are at the final stage.
The above discussions are summarized as follows:
We have considered the string theory in $D$-dimensional
target space with the Liouville-like action,
\be
\frac{1}{8\pi} \int d^2 \xi \sqrt{\hat{g}}
  \left( ( \hat{\nabla} \phi_{q} )^2 \phi_{q}
   + {\cal Q} \hat{R}\phi_{q} \right)
+\frac{1}{8\pi\beta^2} \int d^2 \xi \sqrt{\hat{g}}
  \left( ( \hat{\nabla} \phi_{cl} )^2 \phi_{cl}
   + 2 \hat{R}\phi_{cl} \right),
\ee
where we have defined $\beta=1/\sqrt{\at}$
playing the role of the classical
coupling constant and ${\cal Q}=2 \Delta_{q} \sqrt{\ai}$,
i.e.,
\be
{\cal Q}=\frac{2\Delta_{q}}{\alpha}+\alpha.
\label{Q}
\ee
Further we have the restriction (\ref{cft-condition-2})
which comes from the requirement of
the conformal invariance of the theory;
\be
\frac{D-25}{12}+ \left(\frac{\Delta_{q}}{\alpha}+
\frac{\alpha}{2}\right)^2+
\left(\frac{1-\Delta_{q}}{\beta}\right)^2=0,
\label{rel}
\ee
where the eqs.(\ref{dimension-1}) (\ref{sqrt-1}) are used.
The solution to the eq.(\ref{rel}) is
\be
\alpha=\frac{1}{\sqrt{12}}
\left(
\sqrt{25-D-\frac{12(1-\Delta_q)^2}{\beta^2}}\pm
\sqrt{25-D-\frac{12(1-\Delta_q)^2}{\beta^2}-24\Delta_q}
\right).
\label{alpha}
\ee

It should be noticed here that, when $\Delta_q=1$,
the eqs.(\ref{Q})(\ref{alpha}) reduce to the DDK results,
\be
\alpha\,\longrightarrow\,\gamma=
\frac{\sqrt{25-D}-\sqrt{1-D}}{\sqrt{12}},
\qquad {\cal Q} \longrightarrow\,Q=\sqrt{\frac{25-D}{3}}.
\label{ddk}
\ee
The requirement that $\gamma$ in (\ref{ddk}) should 
be real leads us to the 
severe restriction  $D\leq 1$, known as the \lq\lq D=1" barrier.

On the other hand, our situations are different from DDK case.
We also impose the reality condition on the coupling constant 
$\alpha$, obtaining from (\ref{rel}) or (\ref{alpha});
\be
-\frac{D-25}{12}-\frac{1}{\beta^2}+(\frac{2}{\beta^2}-1)\Delta_{q}-
\frac{\Delta_q^2}{\beta^2}  = 
\frac{\Delta_{q}^2}{\alpha^2}+\frac{\alpha^2}{4}
\geq 2\sqrt{\frac{\Delta_{q}^2}{\alpha^2} \frac{\alpha^2}{4}}
      =   |\Delta_{q}|.
\label{final}
\ee
We have the undetermined constant $\beta$, which
seems to be a coupling constant for the classical sector.
The details of this constant are supposed
to play no physical role in the theory.
Then, we would like to find some continuum
region of values of $\Delta_{q}$,
which must not be dependent on the details of $\beta$.
In general, the regions for $\Delta_q$ satisfying the inequality 
(\ref{final}) depend on $\beta$.  
However, a remarkable case occurs when $\Delta_q<0$. 
If we take $\Delta_{q} > 0$,
we read the relation (\ref{final}) as follows;
\be
D\leq 25-\frac{12(1-\Delta_q)^2}{\beta^2}-24\Delta_q.
\label{final-1}
\ee
The region of $\Delta_{q}$
satisfying the above (\ref{final-1})
is restricted to the $\beta$-dependence by all means.
On the other hand, in case of $\Delta_{q} < 0 $,
the inequality (\ref{final}) becomes as follows;
\be
D\leq 25-\frac{12(1-\Delta_q)^2}{\beta^2}.
\label{final-2}
\ee
In this case, the detailed $\beta$-dependence
of the allowable region fortunately vanishes
as long as $\beta^2 < 0$ and $D\leq 25$
on the relation (\ref{final-2}).
Therefore, we adopt the following conditions
to establish our classical-quantum composite model;
\be
D \leq 25,\quad \quad \Delta_{q} < 0  \quad \mbox{and}
\enskip \beta^2 < 0.
\ee
As a consequence, 
It is clarified that there is a different manner
from well-estsablished
ordinary approach to the 2-d gravity theory,
and we claim that by introducing
the nontrivial classical background of our classical-quantum
composite system,
the target space dimensions $D$
can be beyond ``1'' in the 2-d gravity theory.

Let us make final remarks.
Since we take the condition; $\beta^2<0$,
we then see that the classical object
$\phi_{cl}$ has no equation of motion,
that is, no local information.
Therefore, it instead seems that the classical sector has 
some global information.
Further problems on our model remain to be investigated.\\

\noindent
{\bf Acknowledgments}

We would like to thank Prof. H. Fujisaki
(Rikkyo Univ.) for encouragement.
One of the authors (N.A) is grateful to Professors Yoneya, Kazama
(Univ. of Tokyo) and Saito
(Tokyo Metro. Univ.) for their hospitality.
He also thanks to Y. Sekino for incentive discussions.


\end{document}